\documentclass[prd, aps, 12pt, showpacs,epsfig]{revtex4}
\usepackage{graphicx}
\usepackage{diagbox}
\usepackage{amsmath}
\usepackage{slashbox}
\usepackage{appendix}
\usepackage[perpage,symbol]{footmisc}
\setfnsymbol{wiley}

\begin{document}

\title{ Detecting quark anomalous electroweak couplings at the LHC }

\author{Sheng-Zhi Zhao\footnote{Electronic address: zhao-cz11@mails.tsinghua.edu.cn}, Bin Zhang}
 \affiliation{ Department of Physics,
Tsinghua University, Beijing, 100084, China} \affiliation{Center for
High Energy Physics, Tsinghua University, Beijing, 100084, China}

\begin{abstract}
We study the dimension-6 quark anomalous electroweak couplings in the formulation
of linearly realized effective Lagrangian. We investigate the constraints on these anomalous couplings from the $pp \rightarrow W^+W^-$ process in detail at the LHC. With additional kinematic cuts, we find that the 14 TeV LHC can provide a test of anomalous couplings of $O(0.1-1)\,{\rm TeV}^{-2}$. The $pp \rightarrow ZZ/Z\gamma/\gamma\gamma$ processes can provide a good complement as they are sensitive to those anomalous couplings which do not affect the $pp \rightarrow W^+W^-$ process. Those processes that only contain anomalous triple vertices, like $p p \to W^* \to l \nu_l$ or $p p \to \gamma^* / Z^* \to l^+ l^-$, can improve the sensitivities to $O(0.01-0.1)\,{\rm TeV}^{-2}$. We also study the kinematic differences between different anomalous couplings and discuss the potential of the $\chi^2$-analysis to distinguish them. Finally, we discuss the detection ability of the possible future 100 TeV proton-proton collider(SPPC), and find that the detection sensitivity is improved by about one order of magnitude compared with the LHC.

\end{abstract}


\maketitle

\section{Introduction}

The discovery of the Higgs boson at the Large Hadron Collider (LHC)\cite{Higgsdiscovery} is a great success of the particle physics Standard Model(SM). Meanwhile, there should be some new physics(NP) beyond the SM in order to explain experimental observations such as neutrino oscillation, dark matter required by cosmology and astronomy and the theoretically unnaturalness of SM. Lots of new physics models have been proposed, but no direct evidence for them has been found at the LHC. Current experimental data indicate that the energy scale of new physics $\Lambda_{NP}$ can be higher than several TeV, and new particles would possibly be too heavy to be directly observed at the LHC. As a result, measuring the indirect effects of these models at low energy scales becomes an important way to detect the new physics. The effective Lagrangian has long been an important, model-independent approach for studying the indirect effects of NP. After integrating out heavy degrees of freedom above the cutoff scale $\Lambda$, the leading effects at low energies can be parameterized by the effective interactions between the standard model fields. The coefficients of these interactions, called ``anomalous couplings'', reflect the strength of the new physics effects at low energies.

Many new physics models aim to relieve the fine-tuning problem caused by the SM Higgs sector. These models change the electroweak symmetry breaking(EWSB) Higgs mechanism and cause the anomalous gauge couplings (AGCs) and anomalous Higgs couplings. So AGCs and anomalous Higgs couplings have been well studied. There have been amounts of theoretical studies on AGCs \cite{agctheory} and  experimental measurements at the LEP\cite{agcLEP}, the Tevatron\cite{agcTevatron} and the LHC\cite{agcTevatron, agcATLAS, agcCMS}, and anomalous Higgs couplings are also well studied before\cite{ahcbefore} and after\cite{ahcafter} the Higgs discovery. On the other hand, many new physics models also proposed new gauge interactions or extra fermions. These new sectors would change the couplings between SM fermions and electroweak (EW) gauge bosons directly or indirectly. Different models or model parameters can bring different fermion anomalous couplings deviated from the SM.  There are some model-dependent studies on fermion anomalous couplings, most of them are about top anomalous couplings, like $htt$ type\cite{htt} and $Wtb$ type\cite{wtb}. The effective Lagrangian gives the effective interactions between fermions and electroweak gauge bosons, and measuring these fermion anomalous electroweak couplings is a model-independent method for studying new physics. In our previous paper, we studied how to measure the lepton anomalous electroweak couplings via the $e^+e^- \rightarrow W^+W^-$ process at electron-positron colliders such as the LEP and the ILC\cite{anomalepton}. Similarly there are also quark anomalous electroweak couplings in the effective Lagrangian. We concentrate on how to measure the quark anomalous electroweak couplings via the $pp \rightarrow W^+W^-$ process at the LHC in the present paper. Although the accuracies at hadron colliders are worse than the ILC, the collision energies is much higher. The anomalous couplings will result in higher energy power dependence, hence the high energy $W$-pair production at the LHC can still give sensitive detection  limits on the anomalous couplings. We also expand the study to $pp \to ZZ / Z\gamma / \gamma\gamma$ processes at the LHC. We analyze their measurement capability on quark anomalous electroweak couplings and find that the new processes provide a good complement to the $pp \rightarrow W^+W^-$ process. They are sensitive to those anomalous couplings which are independent to the $pp \rightarrow W^+W^-$ process.

The paper is organized as follows. In Sec.\ref{sec:el} we give the corresponding dimension-6 operators in the effective Lagrangian. In Sec.\ref{sec:ppwwlhc} we study the $pp \rightarrow W^+W^-$ process at the 8 TeV and 14 TeV LHC in details. The anomalous couplings will deviate the cross section of $W$-pair production, and conversely the uncertainty of experimental measurement at the LHC can give constraints on the anomalous couplings. After taking the relative uncertainty of cross section measurement as $\pm 20\%$ at the LHC, the conservative detection limits on the quark anomalous EW couplings are derived. We also introduce some additional kinematic cuts to improve the sensitivities. In Sec.\ref{sec:chi2} we briefly study the kinematic differences between different anomalous couplings and discuss the potential of the $\chi^2$-analysis to distinguish them. In Sec.\ref{sec:comp} $pp \to ZZ / Z\gamma / \gamma\gamma$, $pp \to W^* \to l \nu_l$ and $pp \to \gamma^* / Z^* \to l^+ l^-$  processes at the 14 TeV LHC are taken as complementarities. In Sec.\ref{sec:100TeV} we discuss the detection ability of anomalous couplings at a possible future 100 TeV proton-proton collider(SPPC). In Sec.\ref{sec:summary} we summarize our results.

\section{The Effective Lagrangian}\label{sec:el}

Assuming that there is a fundamental new physics theory above some certain energy scale $\Lambda$, the effective field theory (EFT) is customary to formulate new physics effects  at energies well below $\Lambda$. After integrating out heavy degrees of freedom above $\Lambda$, the leading effects at low energies can be parameterized by the effective
interaction terms that are constructed with the SM fields. In the linearly realizing EFT\cite{EFT-review}, the gauge-invariant high dimensional effective interaction terms are organized in powers of $1/\Lambda$. The effective interaction terms are called effective operators, and their coefficients are called anomalous couplings which reflect the strength of the new physics effects at low energies.

In this paper, we study the quark anomalous EW interactions, so we investigate all the gauge-invariant effective operators that contain quark fields and EW gauge fields. In order to keep the overall dimension consistent in the Lagrangian, any dimension-i effective operator, $\mathcal{O}^{(i)}$, is suppressed by $\frac{1}{\Lambda^{i-4}}$, where $\Lambda$ is the scale of the UV-complete theory. Low energy effects introduced by higher order operators are strongly suppressed, so we only study the leading order dimension-6 effective operators.

We know that CP is not strictly conserved in the SM, and there are CP-odd effective operators in EFT as well. But the CP-odd operators suffer severe experimental constraints, so they are not our concern in this paper. We follow the literature\cite{operator} to construct all these CP-even dimension-6 $SU_c(3)\times SU_W(2)\times U(1)$ gauge invariant operators. They are

\begin{equation}\label{eq:eff-operator}
\begin{split}
\mathcal{O}_{9} & =  i\overline{Q}\gamma_{\mu}W^{\mu\nu}\tensor{D}_{\nu}Q\\
\mathcal{O}_{15} & =  i\overline{Q}\gamma_{\mu}B^{\mu\nu}\tensor{D}_{\nu}Q\\
\mathcal{O}_{17} & =  i\overline{U}\gamma_{\mu}B^{\mu\nu}\tensor{D}_{\nu}U\\
\mathcal{O}_{19} & =  i\overline{D}\gamma_{\mu}B^{\mu\nu}\tensor{D}_{\nu}D\\
\mathcal{O}_{25} & =  \overline{Q}\gamma_{\mu}D_{\nu}W^{\mu\nu}Q\\
\mathcal{O}_{28} & =  \overline{Q}\gamma_{\mu}\partial_{\nu}B^{\mu\nu}Q\\
\mathcal{O}_{29} & =  \overline{U}\gamma_{\mu}\partial_{\nu}B^{\mu\nu}U\\
\mathcal{O}_{30} & =  \overline{D}\gamma_{\mu}\partial_{\nu}B^{\mu\nu}D,
\end{split}
\end{equation}
where $Q$ is the left-handed quark doublet, and $U$($D$) is the right-handed up-type(down-type) quark singlet. $W^i$/ $B$ are $SU_W(2)$ / $U(1)$ gauge fields. $\overline{Q}\tensor{D}_{\nu}Q = \overline{Q} \; D_{\nu}Q - D_{\nu}\overline{Q} \; Q$. The Feynman rules of corresponding anomalous vertices between mass eigenstates after EW symmetry breaking are listed in the appendix. And the effective Lagrangian can be written as

\begin{equation}\label{eq:EL}
\mathcal{L}_{\textrm{eff}} = \mathcal{L}_{\textrm{SM}} + \sum \frac{f_i}{\Lambda^2} \mathcal{O}_i.
\end{equation}

There are always several anomalous couplings affecting one given process. But if the anomalous couplings are small, the deviations from the SM squared amplitudes mainly come from interferences between the SM and anomalous couplings. The interferences between anomalous couplings are negligible, which means that total deviations caused by multi-anomalous couplings is simply the sum of deviations by each coupling. So in this paper we only give results of the single-parameter analysis, as others usually do.

\section{The $pp \rightarrow W^+W^-$ process at the 8 TeV and 14 TeV LHC}\label{sec:ppwwlhc}

The $pp \rightarrow W^+W^-$ process is a good testing process because it contains almost all information of quark EW couplings. The shortage is that it is not associated with $\mathcal{O}_{15}$, $\mathcal{O}_{17}$ and $\mathcal{O}_{19}$ due to the momentum structures of their vertices, just like the case of $\mathcal{O}_{11}$ and $\mathcal{O}_{13}$ in the $e^+ e^- \rightarrow W^+W^-$ process of our previous paper. We take the pure leptonic $W$ decay and the final states are $l^+ \nu_l l^- \bar{\nu}_l$ at hadron colliders, where $l$ stands for $e$ or $\mu$. The final states show up as two opposite-charged leptons with some missing energy on the detectors of the LHC .

\begin{figure}
\centering
\includegraphics[scale=1]{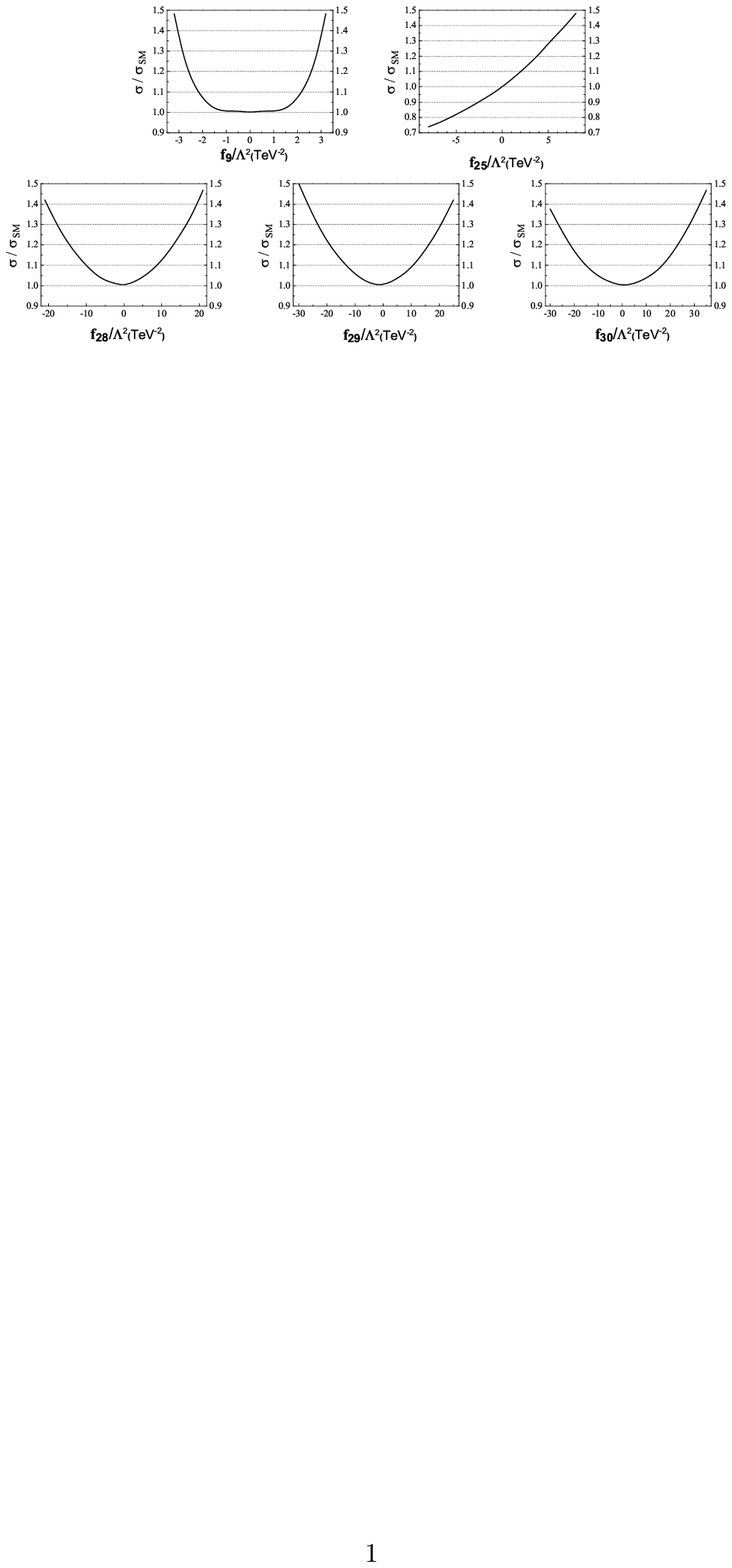}
\caption{The relative deviations from SM cross sections with basic cuts at the 8 TeV LHC.}
\label{fig:8TeV-basic}
\end{figure}

\begin{figure}
\centering
\includegraphics[scale=1]{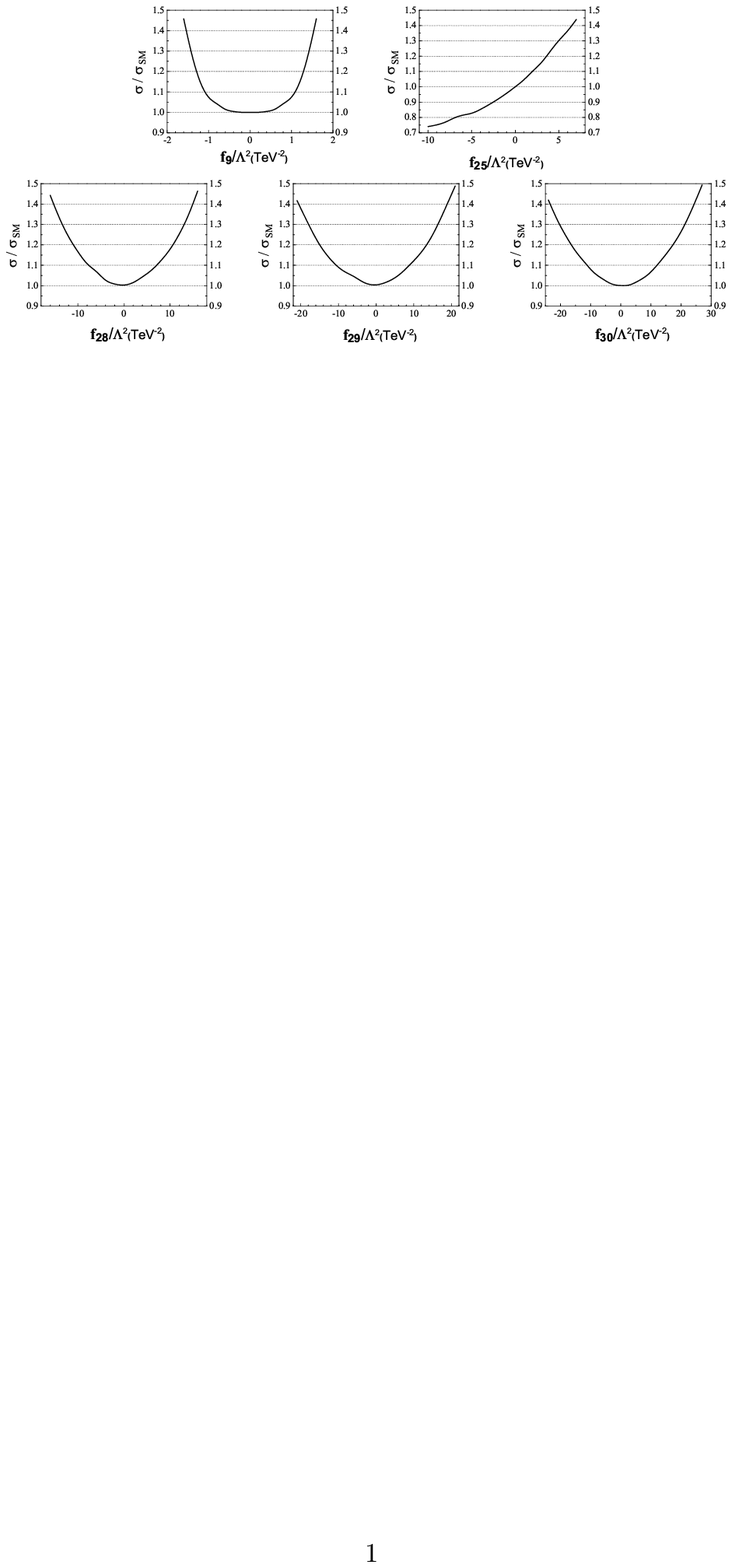}
\caption{The relative deviations from SM cross sections with basic cuts at the 14 TeV LHC.}
\label{fig:14TeV-basic}
\end{figure}

Our first step is to calculate the cross-sections with different anomalous couplings, and derive sensitivity bounds for the anomalous couplings by certain relative deviations from SM predictions. We only calculate the effects of anomalous couplings at tree level, so loop calculations are beyond our study. In order to simulate detectors, we apply the acceptation basic cuts of $|\eta| < 2.5$, $P_T > 10\,\textrm{GeV}$, $\Delta R(l^+ l^-) > 0.4$ on final leptons. Considering that the two opposite-charged leptons may come from a Drell-Yan process or Z-decay, we require $P_T^{\textrm{miss}} > 30\,\textrm{GeV}$ and impose a Z-mass veto as $85\textrm{GeV} < M(l^+ l^-) < 95\,\textrm{GeV}$  if the two final leptons are same-flavor.  After these basic kinematic cuts the remaining cross-section is $\sigma_{SM} = 0.660(1.205)\, \textrm{pb}$ for the 8(14) TeV LHC. Now we apply the single-parameter analysis, i.e.varying the value of one coupling while keeping others zero. Figure \ref{fig:8TeV-basic} and Figure \ref{fig:14TeV-basic} show the relative deviations on cross section caused by different anomalous couplings at the 8 TeV and 14 TeV LHC. $f_9$ and $f_{25}$ affect the t-channel diagram and contribute to the anomalous $qqWW$ vertex so they are different from $f_{28}$, $f_{29}$ and $f_{30}$.

If the experiments at the LHC do not find any obvious deviations from the SM cross section, they can give upper bounds on these anomalous couplings. In our previous paper, we proposed a relative uncertainty as 5\% for the $W$ pair production cross section at the LEP and the ILC. The systematic uncertainties at hadron colliders are larger than electron-positron colliders. In the recent CMS report(the first paper in \cite{agcCMS}), the measured $W^+ W^-$ cross section is $60.1 \pm 4.8$ pb. The relative uncertainty is about 8\%. As we did in our previous paper, we enlarge the uncertainty and conversely propose that the LHC experiments can claim anomalous effects when $|\frac{\sigma-\sigma_{\textrm{SM}}}{\sigma_{\textrm{SM}}}| > 20\%$. This  uncertainty  gives conservative detection limits on $f_i$, and one may have his own uncertainty and derive corresponding detection limits from Figure \ref{fig:8TeV-basic} and Figure \ref{fig:14TeV-basic}. For the 8 TeV LHC our limits are

\begin{equation}\label{eq:bound8}
\begin{split}
-2.6 \textrm{TeV}^{-2} < &f_{9}/\Lambda^{2} < 2.6 \textrm{TeV}^{-2}\\
-5.7 \textrm{TeV}^{-2} < &f_{25}/\Lambda^{2} < 3.8 \textrm{TeV}^{-2}\\
-15 \textrm{TeV}^{-2} < &f_{28}/\Lambda^{2} < 13 \textrm{TeV}^{-2}\\
-19 \textrm{TeV}^{-2} < &f_{29}/\Lambda^{2} < 16 \textrm{TeV}^{-2}\\
-22 \textrm{TeV}^{-2} < &f_{30}/\Lambda^{2} < 24 \textrm{TeV}^{-2},
\end{split}
\end{equation}
and for the 14 TeV LHC the results are

\begin{equation}\label{eq:bound14}
\begin{split}
-1.3 \textrm{TeV}^{-2} < &f_{9}/\Lambda^{2} < 1.3 \textrm{TeV}^{-2}\\
-7.0 \textrm{TeV}^{-2} < &f_{25}/\Lambda^{2} < 3.6 \textrm{TeV}^{-2}\\
-11 \textrm{TeV}^{-2} < &f_{28}/\Lambda^{2} < 11 \textrm{TeV}^{-2}\\
-15 \textrm{TeV}^{-2} < &f_{29}/\Lambda^{2} < 13 \textrm{TeV}^{-2}\\
-17 \textrm{TeV}^{-2} < &f_{30}/\Lambda^{2} < 18 \textrm{TeV}^{-2}.
\end{split}
\end{equation}

Apart from the W-pair production, their decays are also affected by anomalous couplings. But previous measurements tell us that the deviations in the W decay branching fractions from the SM should be less than 2\%. It is a quite small change compared to 20\% deviation in productions, so here we do not take that into consideration. In fact, the measurements of W/Z decay width and branching ratios have high accuracies, and they can also provide constraints on anomalous couplings. We performed analytical analysis of the anomalous leptonic partial width in our previous paper\cite{anomalepton}, and there is no difference for W/Z's hadronic decay.

The main difference between anomalous and SM EW vertices is that the anomalous vertices depend on momentums while the couplings are constants in the SM. It means that the anomalous part becomes more significant when particles carry larger momentum. As we can see from Figure \ref{fig:8TeV-basic} and \ref{fig:14TeV-basic}, the 14 TeV LHC has better anomalous-coupling sensitivities than the 8 TeV LHC. We have used this property to illustrate that the ILC has advantages over the LEP in our previous paper. On the other hand, when we study $pp$ collision we should use parton distribution function (PDF), which means that in a proton-proton collider with certain energy the actual processes have different $E_{CM}$. We know from above discussions that processes with higher $E_{CM}$ tend to be more sensitive to anomalous couplings, so we can improve the sensitivities not only by lifting the collider energy, but also by introducing some kinematic cuts to remove the low-energy parts.

Of course, we can not freely introduce cuts without restriction. We must keep enough number of events for statistic analysis, i.e. make sure that the remaining cross section is above some certain value to really observe deviations if they exist. For $3\sigma$ significance we have

\begin{equation}\label{eq:int1}
\frac{N_S}{\sqrt{N_B+N_S}} = \frac{N-N_{SM}}{\sqrt{N}} = \sqrt{\mathcal{L}_{\textrm{int}}} \frac{\sigma-\sigma_{SM}}{\sqrt{\sigma}} =3,
\end{equation}
where $\mathcal{L}_{\textrm{int}}$ is the integral luminosity.

We require $|\frac{\sigma-\sigma_{\textrm{SM}}}{\sigma_{\textrm{SM}}}| > 0.2$. Taking $\sigma = 1.2 \sigma_{SM}$, we have the needed cross section $\sigma_{SM} = \frac{270}{\mathcal{L}_{\textrm{int}}}$. Run I of the LHC at 7 TeV and 8 TeV has about $\mathcal{L}_{\textrm{int}} = 20 \textrm{fb}^{-1}$, and Run II at 13 TeV and 14 TeV should reach at least $\mathcal{L}_{\textrm{int}} = 300 \textrm{fb}^{-1}$, so the remaining cross section should be no less than several fb.

\begin{figure}
\centering
\includegraphics[scale=1]{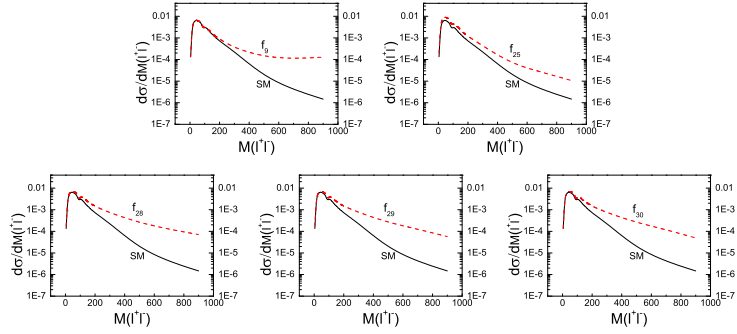}
\caption{$M(l^+ l^-)$ distribution of the cross section at the 8 TeV LHC. The black solid lines and red dash lines indicate SM and anomalous cases, and we set $f_{9}/\Lambda^{2}=3.2 \textrm{TeV}^{-2}$, $f_{25}/\Lambda^{2}=8.2 \textrm{TeV}^{-2}$, $f_{28}/\Lambda^{2}=22 \textrm{TeV}^{-2}$, $f_{29}/\Lambda^{2}=28 \textrm{TeV}^{-2}$ and $f_{30}/\Lambda^{2}=36 \textrm{TeV}^{-2}$ in the five diagrams respectively.}
\label{fig:8TeV-dis-Mll}
\end{figure}

\begin{figure}
\centering
\includegraphics[scale=1]{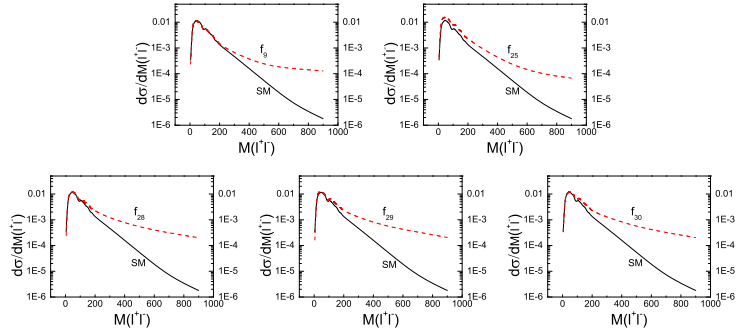}
\caption{$M(l^+ l^-)$ distribution of the cross section at the 14 TeV LHC. The black solid lines and red dash lines indicate SM and anomalous cases, and we set $f_{9}/\Lambda^{2}=1.65 \textrm{TeV}^{-2}$, $f_{25}/\Lambda^{2}=8.0 \textrm{TeV}^{-2}$, $f_{28}/\Lambda^{2}=16 \textrm{TeV}^{-2}$, $f_{29}/\Lambda^{2}=21 \textrm{TeV}^{-2}$ and $f_{30}/\Lambda^{2}=27 \textrm{TeV}^{-2}$ in the five diagrams respectively.}
\label{fig:14TeV-dis-Mll}
\end{figure}

We choose the invariant mass of two final leptons $M(l^+l^-)$ as a kinematic parameter to identify the collision energy. To clearly show the differences in $M(l^+l^-)$ distributions between SM and anomalous cases, we specify particular $f_i$ values to make $\sigma = 1.5\sigma_{\textrm{sm}}$, and draw the distributions in Figure \ref{fig:8TeV-dis-Mll} and Figure \ref{fig:14TeV-dis-Mll}.

It is easy to see that if we only consider the large $M(l^+ l^-)$ range we will get better results. So we introduce additional cuts on $M(l^+ l^-)$ to improve sensitivities, while keeping cross sections above several fb. The results are shown in Table \ref{table:8TeVsummary} and \ref{table:14TeVsummary}.

\begin{table}
\caption{The boundaries of anomalous couplings at the 8 TeV LHC.}\label{table:8TeVsummary}
\begin{tabular}{|c|c|c|c|c|c|c|}
 \tableline
cut & $\sigma_{sm}$ & $\frac{f_{9}}{\Lambda^2}$($\textrm{TeV}^{-2}$) & $\frac{f_{25}}{\Lambda^2}$($\textrm{TeV}^{-2}$) & $\frac{f_{28}}{\Lambda^2}$($\textrm{TeV}^{-2}$) & $\frac{f_{29}}{\Lambda^2}$($\textrm{TeV}^{-2}$) & $\frac{f_{30}}{\Lambda^2}$($\textrm{TeV}^{-2}$)    \\
 \tableline
basic cut & 0.660 pb & (-2.6,2.6) & (-5.7,3.8) & (-15,13) & (-19,16) & (-22,24) \\
 \tableline
$M(l^+ l^-) > 200\textrm{GeV}$ & 51.4 fb & (-1.4,1.4) & (-12,2.6) & (-5.3,5.2) & (-6.2,6.2) & (-7.9,8.6)  \\
  \tableline
\end{tabular}
\end{table}

\begin{table}
\caption{The boundaries of anomalous couplings at the 14 TeV LHC.}\label{table:14TeVsummary}
\begin{tabular}{|c|c|c|c|c|c|c|}
 \tableline
cut & $\sigma_{sm}$ & $\frac{f_{9}}{\Lambda^2}$($\textrm{TeV}^{-2}$) & $\frac{f_{25}}{\Lambda^2}$($\textrm{TeV}^{-2}$) & $\frac{f_{28}}{\Lambda^2}$($\textrm{TeV}^{-2}$) & $\frac{f_{29}}{\Lambda^2}$($\textrm{TeV}^{-2}$) & $\frac{f_{30}}{\Lambda^2}$($\textrm{TeV}^{-2}$)    \\
 \tableline
basic cut & 1.205 pb & (-1.3,1.3) & (-7.0,3.6) & (-11,11) & (-15,13) & (-17,18) \\
 \tableline
$M(l^+ l^-) > 500\textrm{GeV}$ & 6.43 fb & (-0.37,0.37) & (-1.9,1.0) & (-1.3,1.3) & (-1.7,1.5) & (-2.0,2.1)  \\
  \tableline
\end{tabular}
\end{table}

\section{Distinguishing variant anomalous couplings by the $\chi^2$-analysis}\label{sec:chi2}

There have been plenty of studies on anomalous couplings, and their detections or constraints at different colliders. All these studies are mainly based on single-parameter analysis, and one may wonder whether we can tell which anomalous coupling causes the deviation if any signal arises. The single-parameter analysis is just an idealized approach, as all effective operators combine to give the actual result of a typical process. To find out the origin of any anomalous signal, we may choose one specific process(or kinematic variable) which is most sensitive to one anomalous coupling(or a particular combination of the couplings) while insensitive to the others. Then this process, or kinematic variable, will be a good probe for the selected anomalous coupling.

It is easy to figure out whether a given process is a good probe for some anomalous coupling. For example, the $pp \rightarrow W^+W^-$ process is most sensitive to $f_9$ compared to the other four, so it is a good probe for $f_9$. To quantitatively illustrate the validity of some kinematic variable as a probe, we use the idea of $\chi^2$-analysis\cite{chi2} which has been used to distinguish $Z'$ effects from anomalous gauge couplings\cite{Z'-agc}. Even with the same overall cross section, there will be differences between various anomalous couplings which may show up in the kinematic distributions of cross section. Assuming a coupling $f_a$ as benchmark and its value being fixed to give $\sigma=1.2\sigma_{SM}$, we pick some kinematic variable $\kappa$ and divide $\sigma$ into $n$ pieces $\sigma(f_a,i) (i=1,...,n)$ according to its $\kappa$ distribution. Then we vary the value of a different coupling $f_b$ and calculate the kinematic difference as

\begin{equation}\label{eq:chi2}
\begin{split}
\chi^{2}(f_{b},f_{a}) & =  {\sum_{i=1}^n}\left[\frac{N(f_{b},i)-N(f_{a},i)}{\sqrt{N(f_{a},i)}}\right]^{2}\\
 & =  \mathcal{L}_{\textrm{int}}{\sum_{i=1}^n}\left[\frac{(\sigma(f_{b},i)-\sigma(f_{a},i))^{2}}{\sigma(f_{a},i)}\right].
\end{split}
\end{equation}

Scanning all values of $f_b$ we will get $\chi^{2}_{min}(f_{b},f_{a})$, the minimal value of $\chi^{2}(f_{b},f_{a})$. As we know, the general $\chi^{2}$-analysis which compare theoretical predictions with experimental observations  are considered to match if $\chi^2/\textrm{d.o.f} \sim 1$. On the other hand, if $\chi^2/\textrm{d.o.f} \gg 1$, we have poor theoretical predictions. So in our case, as $\chi^2_{min}/\textrm{d.o.f} \gg 1$, we can take $\kappa$ as a good choice to distinguish $f_a$ from $f_b$.(The degrees of freedom is $n-1$ for $n$ pieces and 1 parameter.)

We again consider the kinematic distributions of $\sigma_{pp \rightarrow W^+W^-}$. We set $\mathcal{L}_{\textrm{int}} = 300 \textrm{fb}^{-1}$, and give the results for different $\kappa$ like $M(ll)$, $\Delta\eta(ll)$ or $\phi_{xy}(ll)$ at the 14 TeV LHC. The corresponding $\sigma$ and $f_a$ are already given in Table \ref{table:14TeVsummary}. The results are shown in Table \ref{table:basicchi}-\ref{table:phichi}.

\begin{table}
\caption{Results of $\chi^2_{min} / \textrm{d.o.f}$ with only basic cuts and $\sigma(0 < M(ll) < 1000\textrm{GeV})$ divided into 10 bins.}\label{table:basicchi}
\begin{tabular}{|c|c|c|c|c|c|}
 \tableline
\diagbox[dir=SE]{$f_a$}{$\frac{\chi^2_{min}}{\textrm{d.o.f}}$}{$f_b$} & $f_{9}$ & $f_{25}$ & $f_{28}$ & $f_{29}$ & $f_{30}$\\
 \tableline
$f_{9}$ & $\backslash$ & 1342 & 4289 & 3244 & 4867  \\
 \tableline
$f_{25}$ & 1752 & $\backslash$ & 11480 & 9101 & 12400  \\
 \tableline
$f_{28}$ & 1945 & 3059 & $\backslash$ & 97.32 & 46.59  \\
 \tableline
$f_{29}$ & 1538 & 2627 & 130.0 & $\backslash$ & 125.8  \\
 \tableline
$f_{30}$ & 2077 & 3155 & 45.1 & 105.7 & $\backslash$  \\
 \tableline
 \end{tabular}
\end{table}

\begin{table}
\caption{Results of $\chi^2_{min} / \textrm{d.o.f}$ with $M(ll) > 500\textrm{GeV}$ cut and $\sigma(500 < M(ll) < 1000\textrm{GeV})$ divided into 10 bins.}\label{table:Mllchi}
\begin{tabular}{|c|c|c|c|c|c|}
 \tableline
\diagbox[dir=SE]{$f_a$}{$\frac{\chi^2_{min}}{\textrm{d.o.f}}$}{$f_b$} & $f_{9}$ & $f_{25}$ & $f_{28}$ & $f_{29}$ & $f_{30}$\\
 \tableline
$f_{9}$ & $\backslash$ & 8.569 & 11.43 & 9.249 & 7.095  \\
 \tableline
$f_{25}$ & 6.904 & $\backslash$ & 0.886 & 0.424 & 0.492  \\
 \tableline
$f_{28}$ & 8.829 & 0.835 & $\backslash$ & 0.499 & 0.731  \\
 \tableline
$f_{29}$ & 7.326 & 0.418 & 0.521 & $\backslash$ & 0.722  \\
 \tableline
$f_{30}$ & 5.708 & 0.527 & 0.788 & 0.728 & $\backslash$  \\
 \tableline
 \end{tabular}
\end{table}

\begin{table}
\caption{Results of $\chi^2_{min} / \textrm{d.o.f}$ with $M(ll) > 500\textrm{GeV}$ cut and $\sigma(0 < \Delta\eta(ll) < 5)$ divided into 10 bins.}\label{table:etachi}
\begin{tabular}{|c|c|c|c|c|c|}
 \tableline
\diagbox[dir=SE]{$f_a$}{$\frac{\chi^2_{min}}{\textrm{d.o.f}}$}{$f_b$} & $f_{9}$ & $f_{25}$ & $f_{28}$ & $f_{29}$ & $f_{30}$\\
 \tableline
$f_{9}$ & $\backslash$ & 2.174 & 4.457 & 3.384 & 2.232  \\
 \tableline
$f_{25}$ & 1.990 & $\backslash$ & 1.027 & 0.513 & 0.602  \\
 \tableline
$f_{28}$ & 3.760 & 0.951 & $\backslash$ & 0.552 & 1.861  \\
 \tableline
$f_{29}$ & 2.839 & 0.494 & 0.580 & $\backslash$ & 0.792  \\
 \tableline
$f_{30}$ & 2.040 & 0.587 & 1.965 & 0.832 & $\backslash$  \\
 \tableline
 \end{tabular}
\end{table}

\begin{table}
\caption{Results of $\chi^2_{min} / \textrm{d.o.f}$ with $M(ll) > 500\textrm{GeV}$ cut and $\sigma(2.2 < \phi_{xy}(ll) < 3.2)$ divided into 10 bins.}\label{table:phichi}
\begin{tabular}{|c|c|c|c|c|c|}
 \tableline
\diagbox[dir=SE]{$f_a$}{$\frac{\chi^2_{min}}{\textrm{d.o.f}}$}{$f_b$} & $f_{9}$ & $f_{25}$ & $f_{28}$ & $f_{29}$ & $f_{30}$\\
 \tableline
$f_{9}$ & $\backslash$ & 12.42 & 14.21 & 13.40 & 12.53  \\
 \tableline
$f_{25}$ & 14.71 & $\backslash$ & 1.316 & 0.867 & 1.954  \\
 \tableline
$f_{28}$ & 13.83 & 1.245 & $\backslash$ & 2.234 & 4.524  \\
 \tableline
$f_{29}$ & 14.67 & 0.902 & 2.335 & $\backslash$ & 4.352  \\
 \tableline
$f_{30}$ & 14.92 & 1.574 & 3.422 & 2.828 & $\backslash$  \\
 \tableline
 \end{tabular}
\end{table}

In Table \ref{table:basicchi} all $\chi^2_{min} / \textrm{d.o.f}$ is in orders of magnitude larger than 1, which means that all the five anomalous couplings can be completely distinguished from each other. However, in deriving the results of Table \ref{table:basicchi} we consider the cross section after basic cuts, and we got poor precision of anomalous couplings in this case according to Table \ref{table:14TeVsummary}. We must add additional kinematic cuts to get better precision, but the total cross section is thus reduced and we have smaller $\chi^2_{min} / \textrm{d.o.f}$. From Table \ref{table:Mllchi}-\ref{table:phichi} which are derived after the cut $M(ll) > 500\textrm{GeV}$, we can see that $f_9$ can be distinguished from the other four couplings if we use $M(ll)$ or $\phi_{xy}(ll)$ as a probe, meanwhile $\Delta\eta(ll)$ is invalid. Furthermore, it is hard to recognize $f_{25}$, $f_{28}$, $f_{29}$ and $f_{30}$, as their $\chi^2_{min} / \textrm{d.o.f}$ is just of order one. This is not surprising because the corresponding effective operator $\mathcal{O}_{25}$, $\mathcal{O}_{28}$, $\mathcal{O}_{29}$ and $\mathcal{O}_{30}$ have very similar structures, while $\mathcal{O}_9$ is totally different.

\section{Complementary Processes at the 14 TeV LHC}\label{sec:comp}

To thoroughly study the anomalous couplings, we should take other processes into consideration. For example we get no information of $\mathcal{O}_{15}$, $\mathcal{O}_{17}$ and $\mathcal{O}_{19}$ by studying the $pp \rightarrow W^+W^-$ process. So we need other processes as complementarities. To study $\mathcal{O}_{15}$, $\mathcal{O}_{17}$ and $\mathcal{O}_{19}$ we consider $pp \rightarrow ZZ/Z\gamma/\gamma\gamma$. In these processes, gauge bosons only appear in external lines, so in scattering amplitudes they are replaced with polarization vectors and their momentums are on-shell. We know that $\epsilon(\vec{k}) \cdot \vec{k}= 0$ and $\vec{k} \cdot \vec{k} = m^2$, so in anomalous vertices the parts of $\mathcal{O}_{25}$, $\mathcal{O}_{28}$, $\mathcal{O}_{29}$ and $\mathcal{O}_{30}$ spontaneously cancel to zero or some momentum-independent constant. As a result, we can just consider first four operators. Furthermore, it is known that anomalous triple gauge couplings can also affect the $pp \rightarrow W^+W^-$ process, but if we insist $SU(3) \times SU(2) \times U(1)$, C and P invariance in EW gauge boson interactions, there will be no anomalous couplings of neutral gauge bosons, so $pp \rightarrow ZZ/Z\gamma/\gamma\gamma$ processes are not affected by common AGCs and are thus good choices to distinguish these two kinds of anomalous couplings.

We only consider pure-leptonic Z decays, so the final states are charged leptons and photons. We set basic cuts of $|\eta| < 2.5$, $P_T > 10\textrm{GeV}$, $\Delta R > 0.4$. Assuming that anomalous relative deviations from the SM is within 20\%, the boundaries are given in Table \ref{table:complementarity}.

\begin{table}
\caption{Boundaries for anomalous couplings in complementary processes at the 14 TeV LHC.}\label{table:complementarity}
\begin{tabular}{|c|c|c|c|c|c|}
 \tableline
Process & $\sigma_{sm}$ & $f_{9}$ & $f_{15}$ & $f_{17}$ & $f_{19}$\\
 \tableline
$pp \rightarrow ZZ$ & 18.67fb & (-4.2,4.2) & (-4.1,4.1) & (-4.1,4.1) & (-4.5,4.5)  \\
 \tableline
$pp \rightarrow Z\gamma$ & 2.812pb & (-7.0,7.0) & (-3.2,3.2) & (-3.5,3.5) & (-4.0,4.0)  \\
 \tableline
$pp \rightarrow Z\gamma$($\Delta R(l^+ l^-) < 1$) & 5.647fb & (-1.4,1.4) & (-0.40,0.40) & (-0.68,0.68) & (-0.75,0.75) \\
 \tableline
$pp \rightarrow \gamma\gamma$ & 159.9pb & (-10,10) & (-2.5,2.5) & (-3.0,3.0) & (-3.8,3.8)  \\
 \tableline
$pp \rightarrow \gamma\gamma$($M(\gamma\gamma) > 500\textrm{GeV}$) & 65.93fb & (-0.95,0.95) & (-0.25,0.25) & (-0.28,0.28) & (-0.33,0.33)  \\
 \tableline
 \end{tabular}
\end{table}

\begin{figure}
\centering
\includegraphics[scale=0.5]{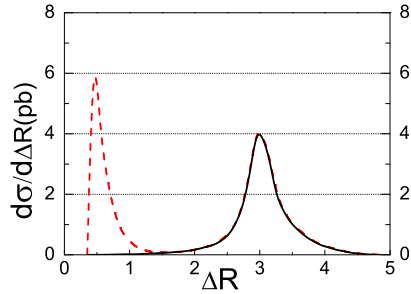}
\caption{$\Delta R(l^+ l^-)$ distribution of $pp \rightarrow Z\gamma$ cross section at the 14 TeV LHC. The black solid lines and red dash lines indicate the SM and anomalous case respectively, and we have set $f_{9}/\Lambda^{2}=8 \textrm{TeV}^{-2}$ for the anomalous case.}
\label{fig:deltar}
\end{figure}

We used a different cut as $\Delta R(l^+ l^-) < 1$ for the $pp \rightarrow Z\gamma$ process. $\Delta R$ is a kinematic variable that describes how separated two final particles are. If a $Z$ boson carries large momentum then its decay products tend to be close at the laboratory frame and $\Delta R(l^+ l^-)$ should be small, hence it does work for our purpose. To see this more clearly, we give an example in Figure \ref{fig:deltar}.

During calculations we find some cancelation between anomalous triple and quartic vertices. For example, if we leave out the anomalous $qqWW$ vertices in the $pp \rightarrow W^+W^-$ process, the sensitivities are improved. So we further consider $p p \to W^* \to l \nu_l$ and $p p \to \gamma^* / Z^* \to l^+ l^-$ processes as they only contain anomalous triple vertices. In these processes, gauge bosons serve as s-channel propagators, and from the discussions in Section.\ref{sec:ppwwlhc} and in the appendix, we can see that they are independent of $f_{9}$, $f_{15}$, $f_{17}$ and $f_{19}$. On the other hand, these processes also contain lepton EW vertices such as $lvW$, $llZ$ and $ll\gamma$, so we should take lepton anomalous EW couplings($f_{24}$, $f_{26}$ and $f_{27}$) into account as well. The experimental precisions of these processes at the LHC are similar to the $pp \rightarrow W^+W^-$ process\cite{CMSll}, so we still take the 20\% limits.

\begin{figure}
\centering
\includegraphics[scale=0.5]{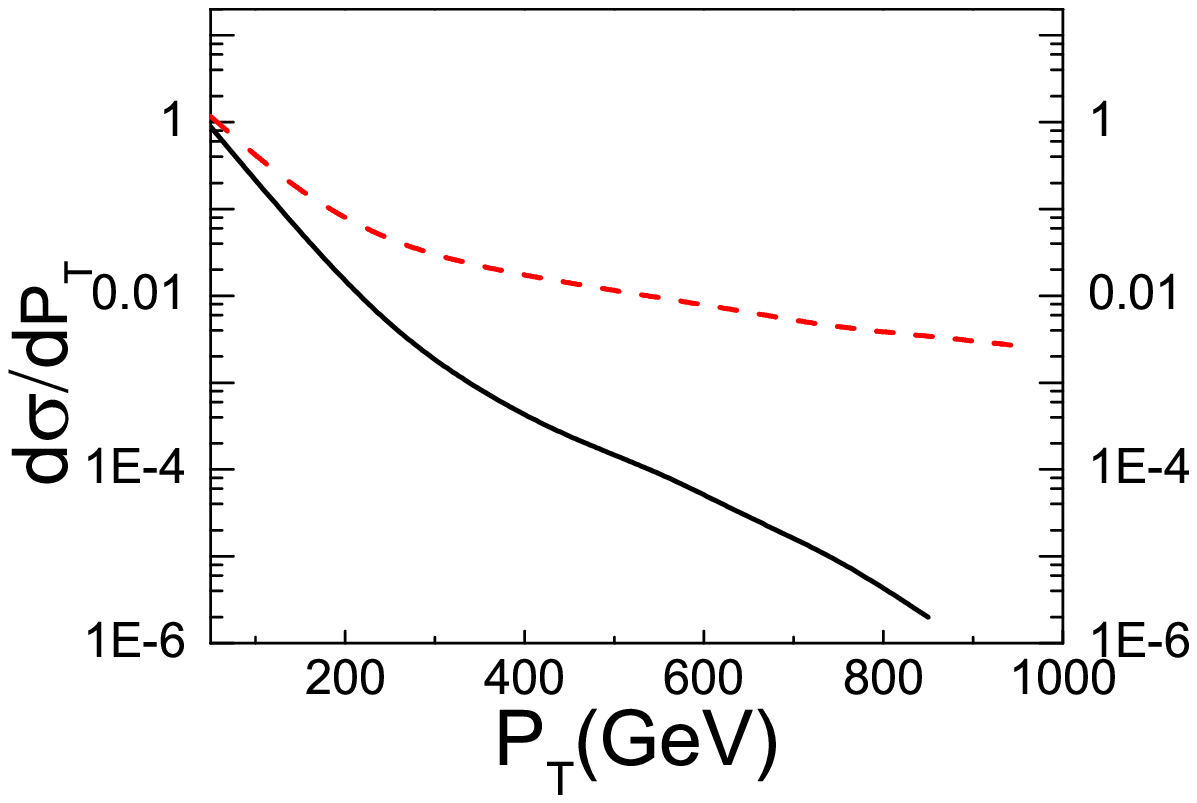}
\caption{$P_T$ distribution of $p p \to l \nu_l$ cross section at the 14 TeV LHC. The black solid lines and red dash lines indicate the SM and anomalous case respectively, and we have set $f_{25}/\Lambda^{2}=4.5 \textrm{TeV}^{-2}$ for the anomalous case.}
\label{fig:pt}
\end{figure}

For the $p p \to l \nu_l$ process the backgrounds are mainly $l + 3 \nu_l$. The lepton transverse momentum $P_T$ can be a useful cut as shown in Figure.\ref{fig:pt}, so we introduce a $P_T^{\textrm{miss}} > 500\textrm{GeV}$ cut to remove the low-energy events. The remaining cross section is $\sigma_{SM} = 13.71 \textrm{fb}$ at the 14 TeV LHC, and the bounds are

\begin{equation}\label{eq:boundlv}
\begin{split}
-0.034 \textrm{TeV}^{-2} < &f_{24}/\Lambda^{2} < 0.030 \textrm{TeV}^{-2}\\
-0.034 \textrm{TeV}^{-2} < &f_{25}/\Lambda^{2} < 0.030 \textrm{TeV}^{-2}.
\end{split}
\end{equation}

For the $p p \to \gamma^* / Z^* \to l^+ l^-$ process the backgrounds are mainly from $t\bar{t}$. To remove the low-energy events we set $M(l^+ l^-) > 1\textrm{TeV}$. The remaining cross section is $\sigma_{SM} = 10.31 \textrm{fb}$ at the 14 TeV LHC, and the bounds are

\begin{equation}\label{eq:boundll}
\begin{split}
-0.063 \textrm{TeV}^{-2} < &f_{24}/\Lambda^{2} < 0.058 \textrm{TeV}^{-2}\\
-0.065 \textrm{TeV}^{-2} < &f_{25}/\Lambda^{2} < 0.059 \textrm{TeV}^{-2}\\
-0.080 \textrm{TeV}^{-2} < &f_{26}/\Lambda^{2} < 0.25 \textrm{TeV}^{-2}\\
-0.066 \textrm{TeV}^{-2} < &f_{27}/\Lambda^{2} < 0.11 \textrm{TeV}^{-2}\\
-0.12 \textrm{TeV}^{-2} < &f_{28}/\Lambda^{2} < 0.046 \textrm{TeV}^{-2}\\
-0.10 \textrm{TeV}^{-2} < &f_{29}/\Lambda^{2} < 0.043 \textrm{TeV}^{-2}\\
-0.089 \textrm{TeV}^{-2} < &f_{30}/\Lambda^{2} < 0.19 \textrm{TeV}^{-2}.
\end{split}
\end{equation}

We finally derive bounds of $O(0.01-0.1)\,{\rm TeV}^{-2}$. For quark anomalous EW couplings the sensitivities are improved by about one or two orders of magnitude compared with the results in Table \ref{table:14TeVsummary}, and the results for lepton anomalous EW couplings are similar to the 1 TeV ILC\cite{anomalepton}.

\section{The $pp \rightarrow W^+W^-$ process at 100 TeV pp colliders}\label{sec:100TeV}

While the LHC is in progress, new colliders with higher energies are under consideration. One of them is some proton-proton collider like the LHC but with a colliding energy of 100 TeV(SPPC)\cite{CEPC-SPPCStudyGroup:2015esa}. We have pointed out that anomalous vertices have positive powers of the momentum, so anomalous couplings are more likely to be detected at 100 TeV colliders than at the LHC. It is notable that, unlike the case of electron-positron colliders where the initial-state energies are definite, in collisions of two high-energy proton beams we will still get a lot of low-energy events due to the parton distribution, so additional cuts that remove the unwanted parts are necessary. The advantage of 100 TeV pp colliders is that we can introduce better kinematic cuts such as $M(ll) > 2 \textrm{TeV}$, and the remaining part will be more sensitive to anomalous couplings than those at the LHC.

We adopt the same basic cuts as at the LHC, but in order to better take advantage of the higher collision energy, we set $P_T^{\textrm{miss}} > 50\textrm{GeV}$. In Table \ref{table:100TeV} we give bounds of anomalous couplings assuming that anomalous relative deviations from the SM is within 20\%. Here we get bounds of $O(0.01) \textrm{TeV}^{-2}$, about 1-2 orders better than the LHC. Though we have only made the comparison in the $pp \rightarrow W^+W^-$ process, it is straightforward to assume that for any process the sensitivities at the 100 TeV pp collider is one order of magnitude or so better than the LHC.

\begin{table}
\caption{Boundaries for anomalous couplings in the $pp \rightarrow W^+W^-$ process at 100 TeV pp colliders.}\label{table:100TeV}
\begin{tabular}{|c|c|c|c|c|c|c|}
 \tableline
cut & $\sigma_{sm}$ & $\frac{f_{9}}{\Lambda^2}$($\textrm{TeV}^{-2}$) & $\frac{f_{25}}{\Lambda^2}$($\textrm{TeV}^{-2}$) & $\frac{f_{28}}{\Lambda^2}$($\textrm{TeV}^{-2}$) & $\frac{f_{29}}{\Lambda^2}$($\textrm{TeV}^{-2}$) & $\frac{f_{30}}{\Lambda^2}$($\textrm{TeV}^{-2}$)    \\
 \tableline
basic cut & 4.46 pb & (-0.07,0.07) & (-5.2,1.9) & (-3.3,3.0) & (-4.2,4.0) & (-4.6,4.8) \\
 \tableline
$M(l^+ l^-) > 2\textrm{TeV}$ & 1.0 fb & (-0.012,0.012) & (-0.078,0.048) & (-0.063,0.056) & (-0.079,0.070) & (-0.093,0.096)  \\
  \tableline
 \end{tabular}
\end{table}

\section{Summary and Outlook}\label{sec:summary}

In this work we have studied the effects of quark anomalous electroweak couplings. We perform a single-parameter analysis and derive constraints on these anomalous couplings assuming that there is no deviation from SM predictions at present and future colliders. We introduce additional kinematic cuts to improve sensitivities, and finally get bounds of $O(0.1-1) \textrm{TeV}^{-2}$ for the LHC and $O(0.01-0.1) \textrm{TeV}^{-2}$ for 100 TeV pp colliders via the $pp \rightarrow W^+W^-$ process. Besides, $pp \rightarrow ZZ/Z\gamma/\gamma\gamma$ processes are also considered in order to study the couplings that do not affect the $pp \rightarrow W^+W^-$ process.

The cancelation between anomalous triple and quartic vertices lead to worse sensitivities, and we have briefly shown how to avoid this cancelation in $p p \to W^* \to l \nu_l$ and $p p \to \gamma^* / Z^* \to l^+ l^-$ processes. Bounds of $O(0.01-0.1)\,{\rm TeV}^{-2}$ are derived both for quark and lepton anomalous EW couplings, which can be similar to the results of the 1 TeV ILC and are one order of magnitude better than the results from the $pp \rightarrow W^+W^-$ process.

If any deviation from SM predictions arises, our analysis can also give hints on the magnitude of anomalous couplings in turn. But if we want to go further in explaining anomalous signals by the effective field theory, the naive single-parameter analysis should be extended. We may search for specific processes that are most sensitive to only one of the couplings while insensitive to the others. For example, the $pp \to W^+ W^-$ process is most sensitive to $f_9$. If we can find proper kinematic variables to further increase its relative sensitivity, then we may take the $pp \to W^+ W^-$ process as a good probe for $f_9$.

{\bf Acknowledgment:} This work of B.~Z. is supported by the
National Science Foundation of China under Grant No. 11135003.

\newpage

\appendix

\section{Feynman Rules for the Anomalous vertices}

We list all the quark anomalous electroweak vertices and corresponding Feynman rules below.\\

\begin{figure}[h]
\centering
\includegraphics[scale=1]{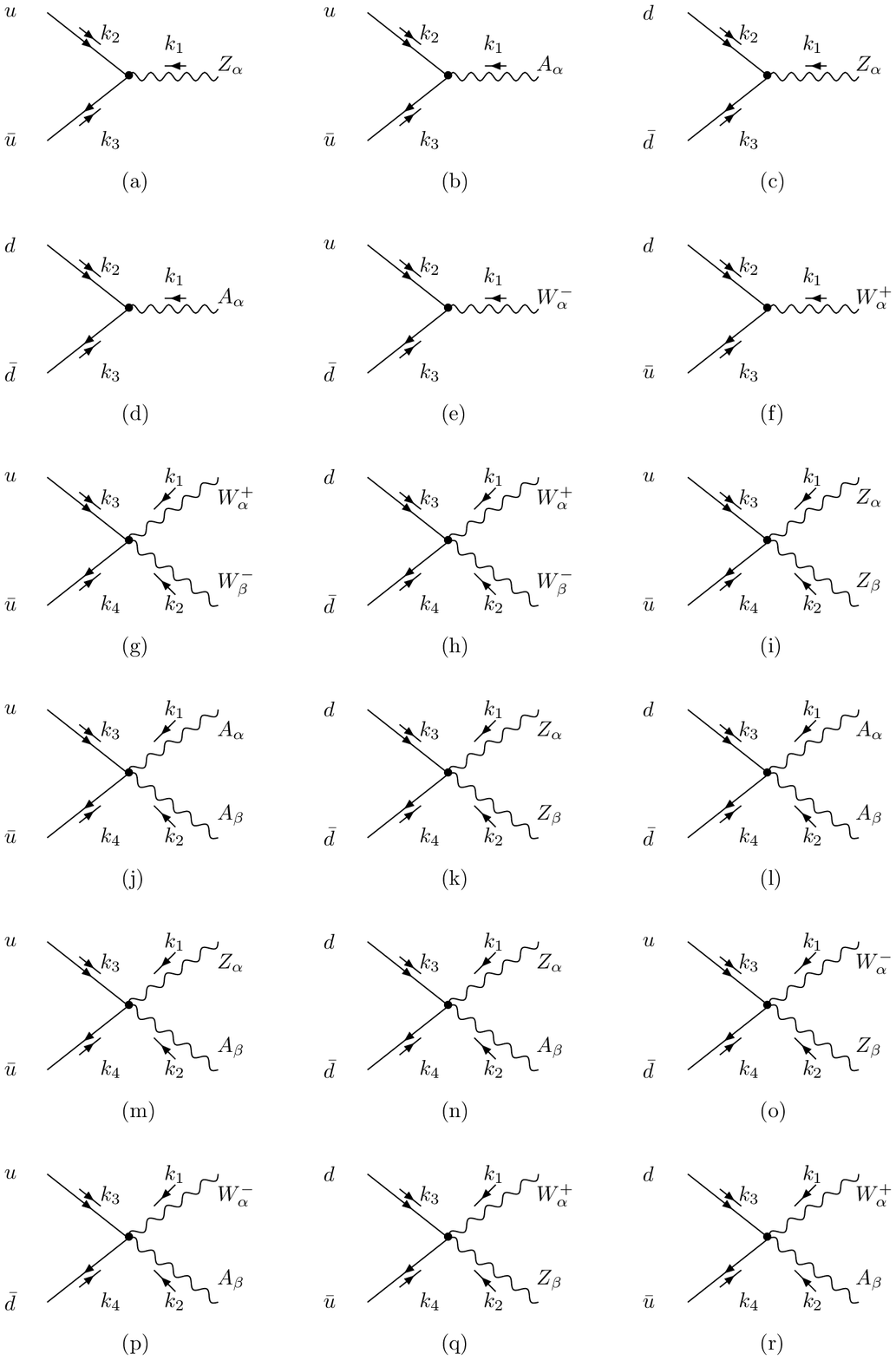}
\end{figure}
%-------------uuZ
\begin{center}
\begin{picture}(350,30)(20,0)
\put(10,30){\makebox(2,1)[l]{$(a)$}}
\put(30,30){\makebox(2,1)[l]{$=$}}
\put(50,30){\makebox(2,1)[l]{$
\frac{1}{\Lambda^2}[k_1{\!\!\!\!\!\slash\;}(k_2-k_3)^\alpha-\gamma^\alpha
k_1\cdot(k_2-k_3)](\frac{c}{2}f_{9}P_L-sf_{15}P_L-sf_{17}P_R)$}}
 \put(50,10){\makebox(2,1)[l]{$
+\frac{i}{\Lambda^2}[k_1{\!\!\!\!\!\slash\;}k_1^\alpha-k_1^2\gamma^\alpha](-\frac{c}{2}f_{25}P_L+sf_{28}P_L+sf_{29}P_R)
$}}
\end{picture}
\end{center}
%-------------uuA
\begin{center}
\begin{picture}(350,30)(20,0)
\put(10,30){\makebox(2,1)[l]{$(b)$}}
\put(30,30){\makebox(2,1)[l]{$=$}}
\put(50,30){\makebox(2,1)[l]{$
\frac{1}{\Lambda^2}[k_1{\!\!\!\!\!\slash\;}(k_2-k_3)^\alpha-\gamma^\alpha
k_1\cdot(k_2-k_3)](\frac{s}{2}f_{9}P_L+cf_{15}P_L+cf_{17}P_R)$}}
 \put(50,10){\makebox(2,1)[l]{$
+\frac{i}{\Lambda^2}[k_1{\!\!\!\!\!\slash\;}k_1^\alpha-k_1^2\gamma^\alpha](-\frac{s}{2}f_{25}P_L-cf_{28}P_L-cf_{29}P_R)
$}}
\end{picture}
\end{center}
%-------------ddZ
\begin{center}
\begin{picture}(350,30)(20,0)
\put(10,30){\makebox(2,1)[l]{$(c)$}}
\put(30,30){\makebox(2,1)[l]{$=$}}
\put(50,30){\makebox(2,1)[l]{$
-\frac{1}{\Lambda^2}[k_1{\!\!\!\!\!\slash\;}(k_2-k_3)^\alpha-\gamma^\alpha
k_1\cdot(k_2-k_3)](\frac{c}{2}f_{9}P_L+sf_{15}P_L+sf_{19}P_R)$}}
 \put(50,10){\makebox(2,1)[l]{$
+\frac{i}{\Lambda^2}[k_1{\!\!\!\!\!\slash\;}k_1^\alpha-k_1^2\gamma^\alpha](\frac{c}{2}f_{25}P_L+sf_{28}P_L+sf_{30}P_R)
$}}
\end{picture}
\end{center}
%-------------ddA
\begin{center}
\begin{picture}(350,30)(20,0)
\put(10,30){\makebox(2,1)[l]{$(d)$}}
\put(30,30){\makebox(2,1)[l]{$=$}}
\put(50,30){\makebox(2,1)[l]{$
-\frac{1}{\Lambda^2}[k_1{\!\!\!\!\!\slash\;}(k_2-k_3)^\alpha-\gamma^\alpha
k_1\cdot(k_2-k_3)](\frac{s}{2}f_{9}P_L-cf_{15}P_L-cf_{19}P_R)$}}
 \put(50,10){\makebox(2,1)[l]{$
+\frac{i}{\Lambda^2}[k_1{\!\!\!\!\!\slash\;}k_1^\alpha-k_1^2\gamma^\alpha](\frac{s}{2}f_{25}P_L-cf_{28}P_L-cf_{30}P_R)
$}}
\end{picture}
\end{center}
%-------------udW
\begin{center}
\begin{picture}(350,30)(20,0)
\put(10,30){\makebox(2,1)[l]{$(e)$}}
\put(30,30){\makebox(2,1)[l]{$=$}}
\put(50,30){\makebox(2,1)[l]{$
\frac{1}{\Lambda^2}[k_1{\!\!\!\!\!\slash\;}(k_2-k_3)^\alpha-\gamma^\alpha
k_1\cdot(k_2-k_3)](\frac{\sqrt{2}}{2}f_{9}P_L) $}}
\put(50,10){\makebox(2,1)[l]{$
-\frac{i}{\Lambda^2}[k_1{\!\!\!\!\!\slash\;}k_1^\alpha-k_1^2\gamma^\alpha](\frac{\sqrt{2}}{2}f_{25}P_L)
$}}
\end{picture}
\end{center}
%-------------duW
\begin{center}
\begin{picture}(350,30)(20,0)
\put(10,30){\makebox(2,1)[l]{$(f)$}}
\put(30,30){\makebox(2,1)[l]{$=$}}
\put(50,30){\makebox(2,1)[l]{$
\frac{1}{\Lambda^2}[k_1{\!\!\!\!\!\slash\;}(k_2-k_3)^\alpha-\gamma^\alpha
k_1\cdot(k_2-k_3)](\frac{\sqrt{2}}{2}f_{9}P_L) $}}
\put(50,10){\makebox(2,1)[l]{$
-\frac{i}{\Lambda^2}[k_1{\!\!\!\!\!\slash\;}k_1^\alpha-k_1^2\gamma^\alpha](\frac{\sqrt{2}}{2}f_{25}P_L)
$}}
\end{picture}
\end{center}
%-------------uuWW
\begin{center}
\begin{picture}(350,30)(20,0)
\put(10,30){\makebox(2,1)[l]{$(g)$}}
\put(30,30){\makebox(2,1)[l]{$=$}}
\put(50,30){\makebox(2,1)[l]{$
\frac{g}{\Lambda^2}[g^{\alpha\beta}(k_1{\!\!\!\!\!\slash\;}+k_2{\!\!\!\!\!\slash\;})
+\gamma^\alpha(k_3-k_1-k_4)^\beta+\gamma^\beta(k_4-k_2-k_3)^\alpha](\frac{1}{2}f_{9}P_L)$}}
 \put(50,10){\makebox(2,1)[l]{$
+\frac{ig}{\Lambda^2}[g^{\alpha\beta}(k_1{\!\!\!\!\!\slash\;}-k_2{\!\!\!\!\!\slash\;})
+\gamma^\alpha(-2k_1-k_2)^\beta+\gamma^\beta(k_1+2k_2)^\alpha](\frac{1}{2}f_{25}P_L)
$}}
\end{picture}
\end{center}
%-------------ddWW
\begin{center}
\begin{picture}(350,30)(20,0)
\put(10,30){\makebox(2,1)[l]{$(h)$}}
\put(30,30){\makebox(2,1)[l]{$=$}}
\put(50,30){\makebox(2,1)[l]{$
\frac{g}{\Lambda^2}[g^{\alpha\beta}(k_1{\!\!\!\!\!\slash\;}+k_2{\!\!\!\!\!\slash\;})
+\gamma^\alpha(k_4-k_1-k_3)^\beta+\gamma^\beta(k_3-k_2-k_4)^\alpha](\frac{1}{2}f_{9}P_L)$}}
 \put(50,10){\makebox(2,1)[l]{$
-\frac{ig}{\Lambda^2}[g^{\alpha\beta}(k_1{\!\!\!\!\!\slash\;}-k_2{\!\!\!\!\!\slash\;})
+\gamma^\alpha(-2k_1-k_2)^\beta+\gamma^\beta(k_1+2k_2)^\alpha](\frac{1}{2}f_{25}P_L)
$}}
\end{picture}
\end{center}
%-------------uuZZ
\begin{center}
\begin{picture}(350,10)(20,0)
\put(10,10){\makebox(2,1)[l]{$(i)$}}
\put(30,10){\makebox(2,1)[l]{$=$}}
\put(50,10){\makebox(2,1)[l]{$
\frac{g}{\Lambda^2}[g^{\alpha\beta}(k_1{\!\!\!\!\!\slash\;}+k_2{\!\!\!\!\!\slash\;})
-\gamma^\alpha k_1^\beta-\gamma^\beta
k_2^\alpha](\frac{4c^2-1}{6}f_{9}P_L-\frac{s(4c^2-1)}{3c}f_{15}P_L+\frac{4s^3}{3c}f_{17}P_R)$}}
\end{picture}
\end{center}
%-------------uuAA
\begin{center}
\begin{picture}(350,10)(20,0)
\put(10,10){\makebox(2,1)[l]{$(j)$}}
\put(30,10){\makebox(2,1)[l]{$=$}}
\put(50,10){\makebox(2,1)[l]{$
\frac{g}{\Lambda^2}[g^{\alpha\beta}(k_1{\!\!\!\!\!\slash\;}+k_2{\!\!\!\!\!\slash\;})
-\gamma^\alpha k_1^\beta-\gamma^\beta
k_2^\alpha](\frac{2s^2}{3}f_{9}P_L+\frac{4cs}{3}f_{15}P_L+\frac{4cs}{3}f_{17}P_R)$}}
\end{picture}
\end{center}
%-------------ddZZ
\begin{center}
\begin{picture}(350,10)(20,0)
\put(10,10){\makebox(2,1)[l]{$(k)$}}
\put(30,10){\makebox(2,1)[l]{$=$}}
\put(50,10){\makebox(2,1)[l]{$
\frac{g}{\Lambda^2}[g^{\alpha\beta}(k_1{\!\!\!\!\!\slash\;}+k_2{\!\!\!\!\!\slash\;})
-\gamma^\alpha k_1^\beta-\gamma^\beta
k_2^\alpha](\frac{2c^2+1}{6}f_{9}P_L+\frac{s(2c^2+1)}{3c}f_{15}P_L-\frac{2s^3}{3c}f_{19}P_R)$}}
\end{picture}
\end{center}
%-------------ddAA
\begin{center}
\begin{picture}(350,10)(20,0)
\put(10,10){\makebox(2,1)[l]{$(l)$}}
\put(30,10){\makebox(2,1)[l]{$=$}}
\put(50,10){\makebox(2,1)[l]{$
\frac{g}{\Lambda^2}[g^{\alpha\beta}(k_1{\!\!\!\!\!\slash\;}+k_2{\!\!\!\!\!\slash\;})
-\gamma^\alpha k_1^\beta-\gamma^\beta
k_2^\alpha](\frac{s^2}{3}f_{9}P_L-\frac{2cs}{3}f_{15}P_L-\frac{2cs}{3}f_{19}P_R)$}}
\end{picture}
\end{center}
%-------------uuZA
\begin{center}
\begin{picture}(350,50)(20,0)
\put(10,50){\makebox(2,1)[l]{$(m)$}}
\put(30,50){\makebox(2,1)[l]{$=$}}
\put(50,50){\makebox(2,1)[l]{$
\frac{g}{\Lambda^2}[g^{\alpha\beta}k_1{\!\!\!\!\!\slash\;}
-\gamma^\alpha k_1^\beta](\frac{2cs}{3}f_{9}P_L-\frac{4s^2}{3}f_{15}P_L)$}}
\put(50,30){\makebox(2,1)[l]{$
+\frac{g}{\Lambda^2}[g^{\alpha\beta}k_2{\!\!\!\!\!\slash\;}
-\gamma^\beta k_2^\alpha](\frac{s(4c^2-1)}{6c}f_{9}P_L+\frac{4c^2-1}{3}f_{15}P_L)$}}
\put(50,10){\makebox(2,1)[l]{$
-\frac{g}{\Lambda^2}[g^{\alpha\beta}(k_1{\!\!\!\!\!\slash\;}+k_2{\!\!\!\!\!\slash\;})
-\gamma^\alpha k_1^\beta-\gamma^\beta
k_2^\alpha](\frac{4s^2}{3}f_{17}P_R)$}}
\end{picture}
\end{center}
%-------------ddZA
\begin{center}
\begin{picture}(350,50)(20,0)
\put(10,50){\makebox(2,1)[l]{$(n)$}}
\put(30,50){\makebox(2,1)[l]{$=$}}
\put(50,50){\makebox(2,1)[l]{$
\frac{g}{\Lambda^2}[g^{\alpha\beta}k_1{\!\!\!\!\!\slash\;}
-\gamma^\alpha k_1^\beta](\frac{cs}{3}f_{9}P_L+\frac{2s^2}{3}f_{15}P_L)$}}
\put(50,30){\makebox(2,1)[l]{$
+\frac{g}{\Lambda^2}[g^{\alpha\beta}k_2{\!\!\!\!\!\slash\;}
-\gamma^\beta k_2^\alpha](\frac{s(2c^2+1)}{6c}f_{9}P_L-\frac{2c^2+1}{3}f_{15}P_L)$}}
\put(50,10){\makebox(2,1)[l]{$
+\frac{g}{\Lambda^2}[g^{\alpha\beta}(k_1{\!\!\!\!\!\slash\;}+k_2{\!\!\!\!\!\slash\;})
-\gamma^\alpha k_1^\beta-\gamma^\beta
k_2^\alpha](\frac{2s^2}{3}f_{19}P_R)$}}
\end{picture}
\end{center}
%-------------udWZ
\begin{center}
\begin{picture}(350,70)(20,0)
\put(10,70){\makebox(2,1)[l]{$(o)$}}
\put(30,70){\makebox(2,1)[l]{$=$}}
\put(50,70){\makebox(2,1)[l]{$
\frac{g}{\Lambda^2}[\gamma^\alpha(k_3-k_4)^\beta-\gamma^\beta(k_3-k_4)^\alpha](\frac{\sqrt{2}}{2}cf_{9}P_L)$}}
\put(50,50){\makebox(2,1)[l]{$
-\frac{g}{\Lambda^2}[g^{\alpha\beta}k_1{\!\!\!\!\!\slash\;}-\gamma^\alpha
k_1^\beta](\frac{\sqrt{2}s^2}{6c}f_{9}P_L)$}}
\put(50,30){\makebox(2,1)[l]{$
-\frac{g}{\Lambda^2}[g^{\alpha\beta}k_2{\!\!\!\!\!\slash\;}-\gamma^\beta
k_2^\alpha](\sqrt{2}sf_{15}P_L)$}}
 \put(50,10){\makebox(2,1)[l]{$
+\frac{ig}{\Lambda^2}[g^{\alpha\beta}(k_1{\!\!\!\!\!\slash\;}-k_2{\!\!\!\!\!\slash\;})-\gamma^\alpha
(2k_1+k_2)^\beta+\gamma^\beta (k_1+2k_2)^\alpha](\frac{\sqrt{2}c}{2}f_{25}P_L)$}}
\end{picture}
\end{center}
%-------------udWA
\begin{center}
\begin{picture}(350,70)(20,0)
\put(10,70){\makebox(2,1)[l]{$(p)$}}
\put(30,70){\makebox(2,1)[l]{$=$}}
\put(50,70){\makebox(2,1)[l]{$
\frac{g}{\Lambda^2}[\gamma^\alpha(k_3-k_4)^\beta-\gamma^\beta(k_3-k_4)^\alpha](\frac{\sqrt{2}}{2}sf_{9}P_L)$}}
\put(50,50){\makebox(2,1)[l]{$
+\frac{g}{\Lambda^2}[g^{\alpha\beta}k_1{\!\!\!\!\!\slash\;}-\gamma^\alpha
k_1^\beta](\frac{\sqrt{2}s}{6}f_{9}P_L)$}}
\put(50,30){\makebox(2,1)[l]{$
+\frac{g}{\Lambda^2}[g^{\alpha\beta}k_2{\!\!\!\!\!\slash\;}-\gamma^\beta
k_2^\alpha](\sqrt{2}cf_{15}P_L)$}}
 \put(50,10){\makebox(2,1)[l]{$
+\frac{ig}{\Lambda^2}[g^{\alpha\beta}(k_1{\!\!\!\!\!\slash\;}-k_2{\!\!\!\!\!\slash\;})-\gamma^\alpha
(2k_1+k_2)^\beta+\gamma^\beta (k_1+2k_2)^\alpha](\frac{\sqrt{2}s}{2}f_{25}P_L)$}}
\end{picture}
\end{center}
%-------------duWZ
\begin{center}
\begin{picture}(350,70)(20,0)
\put(10,70){\makebox(2,1)[l]{$(q)$}}
\put(30,70){\makebox(2,1)[l]{$=$}}
\put(50,70){\makebox(2,1)[l]{$
-\frac{g}{\Lambda^2}[\gamma^\alpha(k_3-k_4)^\beta-\gamma^\beta(k_3-k_4)^\alpha](\frac{\sqrt{2}}{2}cf_{9}P_L)$}}
\put(50,50){\makebox(2,1)[l]{$
-\frac{g}{\Lambda^2}[g^{\alpha\beta}k_1{\!\!\!\!\!\slash\;}-\gamma^\alpha
k_1^\beta](\frac{\sqrt{2}s^2}{6c}f_{9}P_L)$}}
\put(50,30){\makebox(2,1)[l]{$
-\frac{g}{\Lambda^2}[g^{\alpha\beta}k_2{\!\!\!\!\!\slash\;}-\gamma^\beta
k_2^\alpha](\sqrt{2}sf_{15}P_L)$}}
 \put(50,10){\makebox(2,1)[l]{$
-\frac{ig}{\Lambda^2}[g^{\alpha\beta}(k_1{\!\!\!\!\!\slash\;}-k_2{\!\!\!\!\!\slash\;})-\gamma^\alpha
(2k_1+k_2)^\beta+\gamma^\beta (k_1+2k_2)^\alpha](\frac{\sqrt{2}c}{2}f_{25}P_L)$}}
\end{picture}
\end{center}
%-------------duWA
\begin{center}
\begin{picture}(350,70)(20,0)
\put(10,70){\makebox(2,1)[l]{$(r)$}}
\put(30,70){\makebox(2,1)[l]{$=$}}
\put(50,70){\makebox(2,1)[l]{$
-\frac{g}{\Lambda^2}[\gamma^\alpha(k_3-k_4)^\beta-\gamma^\beta(k_3-k_4)^\alpha](\frac{\sqrt{2}}{2}sf_{9}P_L)$}}
\put(50,50){\makebox(2,1)[l]{$
+\frac{g}{\Lambda^2}[g^{\alpha\beta}k_1{\!\!\!\!\!\slash\;}-\gamma^\alpha
k_1^\beta](\frac{\sqrt{2}s}{6}f_{9}P_L)$}}
\put(50,30){\makebox(2,1)[l]{$
+\frac{g}{\Lambda^2}[g^{\alpha\beta}k_2{\!\!\!\!\!\slash\;}-\gamma^\beta
k_2^\alpha](\sqrt{2}cf_{15}P_L)$}}
 \put(50,10){\makebox(2,1)[l]{$
-\frac{ig}{\Lambda^2}[g^{\alpha\beta}(k_1{\!\!\!\!\!\slash\;}-k_2{\!\!\!\!\!\slash\;})-\gamma^\alpha
(2k_1+k_2)^\beta+\gamma^\beta (k_1+2k_2)^\alpha](\frac{\sqrt{2}s}{2}f_{25}P_L)$}}
\end{picture}
\end{center}

In s-channel diagrams of the $pp \to W^+ W^-$ process, we have $k_1 = -(k_2+k_3) = (E_{cm},0,0,0), \; k_2^2 = k_3^2 = 0$ in the center-of-mass frame where $k_1$, $k_2$ and $k_3$ are the momentum of the propagator, the initial-state quark and the initial-state anti-quark respectively. So the s-channel diagrams are independent of $f_9$, $f_{15}$, $f_{17}$ and $f_{19}$ as can be seen from the Feynman rules. $f_9$ appears in the anomalous $qqW$ and $qqWW$ vertex so it still affects the $pp \to W^+ W^-$ process.
\end{document}